# Controlled tuning of whispering gallery modes of GaN/InGaN microdisk cavities

Igor Aharonovich[1], Nan Niu[1], Fabian Rol[1], Kasey J Russell[1], Alexander Woolf[1], Haitham A. R. El-Ella[2], Menno J. Kappers[2], Rachel A Oliver[2] and Evelyn L. Hu[1]

[1]School of Engineering and Applied Sciences, Harvard University, Cambridge, MA, 02138, USA
[2]Department of Materials Science and Metallurgy, University of Cambridge, Pembroke Street, Cambridge CB2 3QZ, United Kingdom

Email: igor@seas.harvard.edu

**Abstract**

Controlled tuning of the whispering gallery modes of GaN/InGaN $\mu$-disk cavities is demonstrated. The whispering gallery mode (WGM) tuning is achieved at room temperature by immersing the $\mu$-disks in water and irradiating with ultraviolet (UV) laser. The tuning rate can be controlled by varying the laser excitation power, with a nanometer precision accessible at low excitation power (~ several µW). The selective oxidation mechanism is proposed to explain the results and supported by theoretical analysis. The tuning of WGMs in GaN/InGaN $\mu$-disk cavities may have important implication in cavity quantum electrodynamics and the development of efficient light emitting devices.

Optical microcavities are important components in studying solid state cavity quantum electrodynamics (cQED), developing low threshold light emitting devices and ultra-sensitive sensors[1-4]. III-nitride materials are particularly interesting in this respect due their high potential in the realization of ultraviolet and visible micro- and nano-scale light sources operating at room temperature. In addition, the strong exciton binding energy of these high bandgap materials makes them attractive candidates for the study of polariton dynamics and polariton lasing[5]. Recent years have seen progress in the formation of III-nitride based cavities that incorporate quantum well (QW) or quantum dot (QD) active layers[6-10]. Nevertheless, an outstanding challenge in engineering III nitride $\mu$–disk cavity systems is the tuning of the cavity modes into resonance with the emission lines of embedded QDs. Such tuning is essential to demonstrate efficient coupling interaction between the cavity mode and the emitter.

A variety of methods have been employed to tune cavity modes in other semiconductor materials (e.g. GaAs) and are mostly based on the modification of the local refractive index of the cavity. Tuning of photonic crystal cavities (PCCs) using gas condensation[11] has been previously demonstrated. In a different approach, deposition of thin photosensitive layer on top of the cavity (such as chalcogenide[12] or spiropyran[13]) is performed. Upon laser irradiation, the refractive index of the layers is modified, resulting in a tuning of the cavity modes. Alternatively, chemical digital etching which slightly modifies the cavity dimensions was also demonstrated.[14] Since there is a different temperature dependence of the QD emission energies and that of the optical modes, change of temperature has been used to effect tuning.[15] Better controlled local tuning has been achieved by employing heating pads positioned in a close proximity with the PCC.[16]

In this paper we demonstrate tuning of the InGaN/GaN $\mu$-disks WGMs through a selective, *in situ* photo-enhanced process. The tuning is achieved by immersing the $\mu$-disks in de-ionized water and exciting the embedded QDs with a UV laser (360 or 380 nm) excitation. The tuning of the WGMs continues as long as the laser is switched on and focused on the disk. In our approach, only the probed $\mu$-disks undergo the tuning while other components on the chip remain unaffected. By measuring cavity luminescence during tuning, we are able to continuously monitor the spectral location of the modes and dynamically tune them.

The structure of the $\mu$-disks consists of an InGaN QD (~20% In) layer sandwiched between two layers of GaN. A detailed description of the fabrication of these cavities and their general characteristics will be published elsewhere[17]. Briefly, the disk structure contains InGaN QDs clad by GaN layers, grown on a sapphire substrate. The growth of the InGaN QDs has been described previously[18]. 500 nm of $SiO_2$ was deposited over the GaN-based structure; the entire sample was bonded onto a secondary sapphire substrate, and the original sapphire was removed by a laser lift-off process. The overlying material, extraneous to the disk structure was removed by a dry etch process and mechanical polishing. The $\mu$-disks were patterned using contact lithography and a post formed by undercutting the SiO$_2$ in hydrofluoric acid. Figure 1a shows a scanning electron microscope (SEM) of the $\mu$-disk. The inset shows a schematic of the disk, displaying InGaN QD layer, ~ 2.5 nm thick, sandwiched between two GaN layers (60 nm each). The diameter of the disks is ~ 3 µm and the thickness is ~ 120 nm. The thin and smooth $\mu$-disks support the propagation of WGMs in the periphery of the disks.

Optical properties of the $\mu$-disks were assessed using a frequency-doubled titanium sapphire laser emitting at 360 nm, incident on the sample through a long working distance objective (×100, numerical aperture (NA)=0.5). The diameter of the beam is approximately 500 nm. The emission from the $\mu$-disks was collected through the same objective and directed into a spectrometer. Figure 1b shows a photoluminescence (PL) spectrum recorded at room temperature from the $\mu$-disk. Modes with quality factor (Q) as high as 3500 were measured.

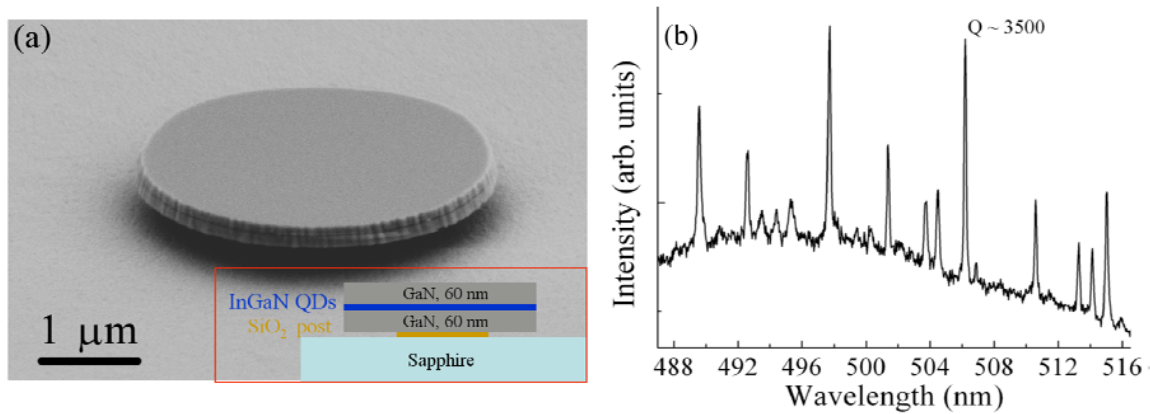

Fig. 1. (color online) (a) SEM image of the InGaN/GaN $\mu$–disk. The inset shows the structure of the disk. (b) PL spectrum recorded from the disk showing the WGMs. Modes with Qs as high as 3500 were measured.

To tune the WGMs, the disks were immersed in water within a small cell. The cell consists of a hollow rubber ring (1 mm height) in between two cover glass slides. The sample was positioned on the bottom cover slide, in the middle of the ring, and imaged using the objective through the top cover slide. Figures 2 (a,c) shows the tuning of the cavity WGMs, achieved by continuous irradiation of the disk with two different excitation powers of 0.2 mW (a) and 0.9 mW (c) using the 360 nm excitation wavelength.

It is noticeable that high excitation powers cause faster WGMs shifts, while lower excitation power results in a more moderate shift.

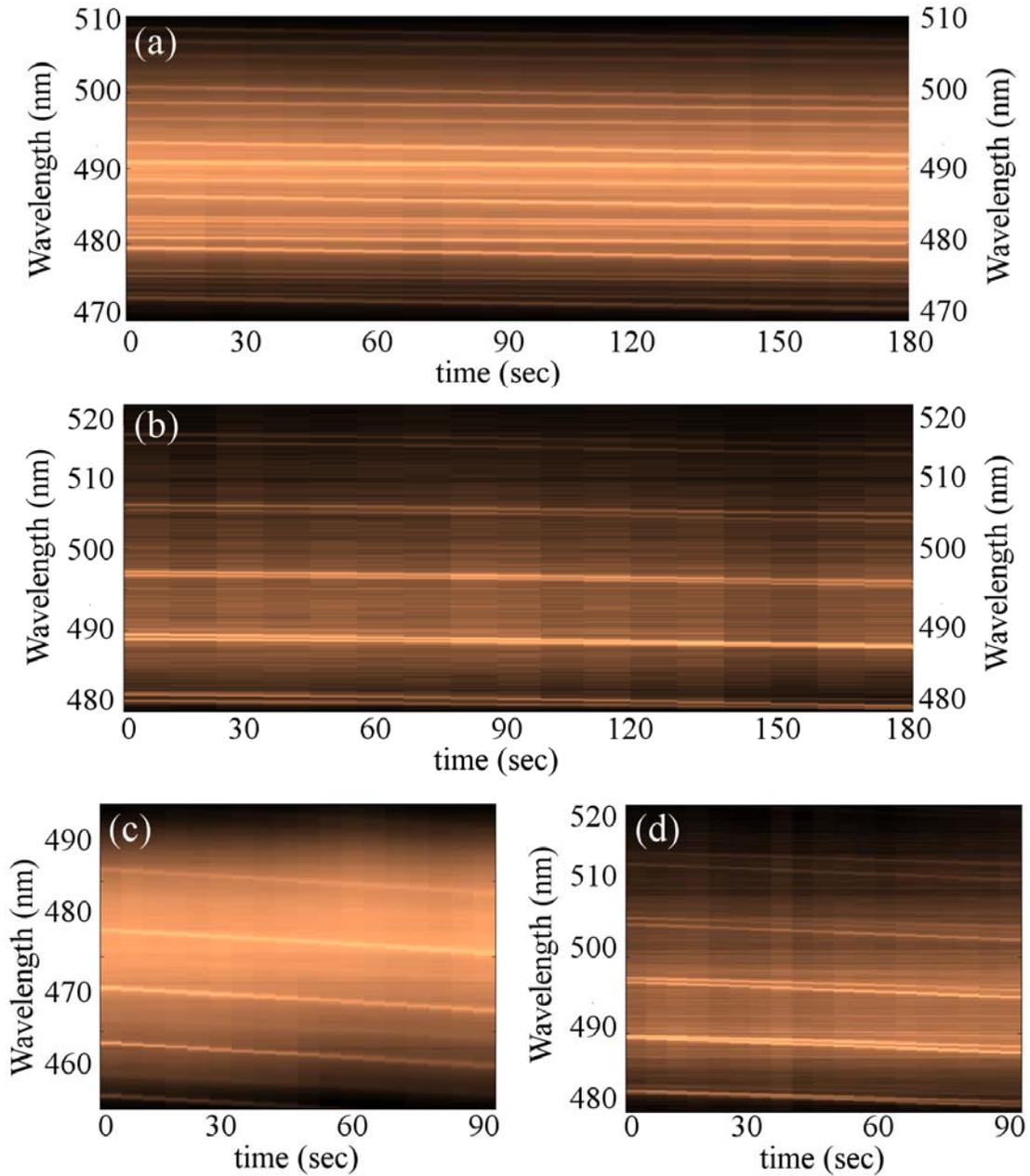

Fig. 2. (color online) Mode tuning of InGaN/GaN $\mu$–disks immersed in water using (a) 360 nm excitation wavelength with 0.2 mW and (c) 0.9 mW excitation power. (b) 380 nm excitation wavelength with 0.3 mW and (d) 1 mW excitation power. The bright lines represent the PL intensity of the different modes.

We repeated the same tuning operation using a 380 nm excitation wavelength, which corresponds to an energy below the bandgap of the GaN. Figures 2 (b,d) show the results of the mode tuning employing 0.3 mW and 1 mW of excitation power. Comparing the two excitation wavelengths, it is clearly seen that tuning using the 380 nm excitation is considerably slower than that observed for the 360 nm excitation. Over the same period of time at excitation powers of ~1 mW, using 380 nm excitation, the modes shifted only by 1.9 nm, while shifts of up to 3.5 nm were observed under 360 nm excitation. The tuning was even slower employing low excitation powers (Fig. 2b) and was not observed when the power dropped to ~50 µW. These data suggest the possibility of precise control of the tuning of the WGMs using low excitation power. Finally, we note that the mode shift is permanent and remains even after the water is removed and the disks are dried in air.

Figures 3 shows two PL spectra recorded dynamically during the tuning after 5 sec (black curves) and 100 sec (red curves) using (a) 360 nm and (b) 380 nm excitation, respectively. A comparison of these spectra reveals a degradation of $Q$ for the 360 nm tuning process, which is minor for the 380 nm tuning, even for a comparable change in wavelength.

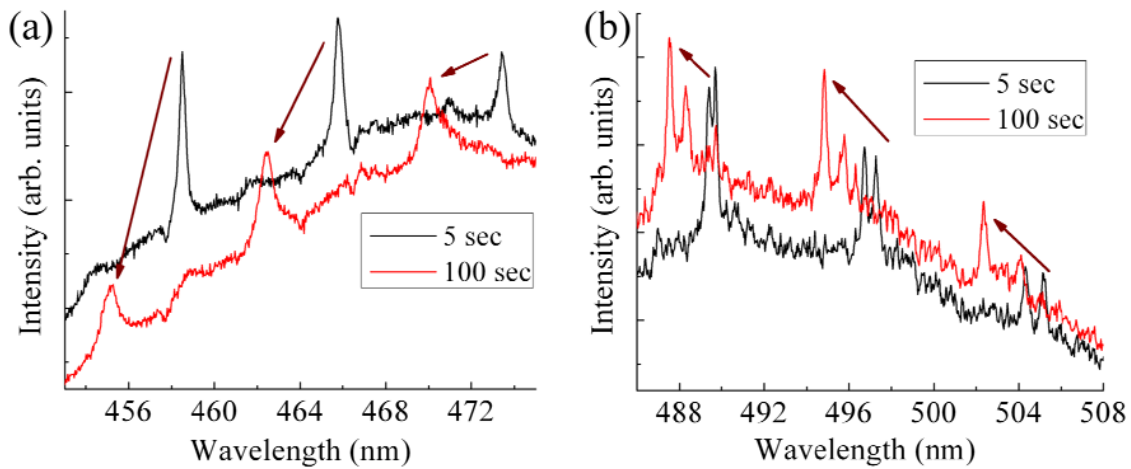

Fig. 3. (color online) PL spectrum recorded after 5 sec (black curve) and 100 sec (red curve) during the tuning of the WGMs using (a) 360 nm and (b) 380 excitation wavelengths. The dark red arrows indicate the mode shift.

If we change the ambient liquid to a non-polar, non-ionic solvent, toluene, we do not observe any shift of the optical modes of the immersed $\mu$–disks, even after 20 minutes of continuous irradiation at 80 µW power with 360 nm excitation laser.

We believe that the tuning mechanism results from a process similar to that seen in Photo Electro Chemical (PEC) etching of GaN[19], in which photo-generated holes enhance oxidation of the GaN, and where the oxide is subsequently dissolved in the electrolyte (water in our case). The consistent blue shift of the modes (several nanometers) with increased tuning power and time suggests a reduction in the size of the $\mu$–disks. We note that typical PEC etching conditions employed for the smoothest-quality etching of GaN incorporates a dilute, 0.004 M HCl solution as the electrolyte[7]. In addition, there has been a previous report of photoelectrochemical oxidation of n-GaN in water, under an applied bias > 2V[20]. In that case, oxides of several 100 nm were formed, but no oxides were observed for biases less than 2V. We believe that in our case, photo-oxidation takes place, with a thin oxide that is subsequently dissolved in solution, resulting in a precise etch removal of the disk material that produces a shift in the cavity modes.

The mechanism of the *in situ* PEC etch-tuning also accords well with the wavelength-dependence of the tuning. For excitation at 380 nm, below the bandgap of the GaN, electron hole pairs will only be generated within the InGaN QDs layer. At low incident powers (~ 50 µW), the photo-generated carriers would be confined in the InGaN region, and thus no PEC etching is observed. At higher powers, band-filling results, the electrons and holes may no longer be confined by the InGaN region, and will drift and diffuse to the $\mu$–disk surfaces, under the influence of the built-in electric field (Fig S1). We believe that, given the direction of the internal fields, photo-oxidation/etching would take place predominantly on the top surface (N-face), although this would need to be further verified. Excitation at 360 nm will generate electron-hole pairs throughout the volume of the disk. With high photo-generated electrons and holes, and no preferential removal of the electrons, we would expect considerable electron-hole recombination. This would limit the efficacy of the holes in driving photo-oxidation and etching, *except* at the periphery of the disk, where the holes are within a diffusion length of the surface. This should, result in a fairly isotropic photo-oxidation/etching of the periphery of the disk, as well as the surfaces.

COMSOL software was used to simulate the shift in wavelength of the modes with respect to change in dimensions of the $\mu$–disk (Fig. S2). A change of 10 nm in the radius of the $\mu$–disk produces only a 1 nm shift in the mode wavelength, while a 10 nm change in the thickness of the $\mu$–disk, results in approximately 7 nm wavelength shift of the WGMs. Thus an 'isotropic' etch of the disk, with a reduction in both the thickness and diameter of the disk (as would apply to the 360 nm tuning excitation) would exhibit a mode shift very nearly equal to an etch that affected only the thickness of $\mu$–disk (characteristic of the 380 nm tuning excitation). The WGMs are extremely sensitive to both changes in the disk diameter and to the quality of the etched sidewalls. Thus, for a similar observed mode shift, the isotropic etch process associated with the 360 nm excitation could have a greater impact on the dimensions and quality of the $\mu$–disk sidewalls. Indeed, a larger degradation of Q seen for the 360 nm tuning process, compared with that undertaken using excitation at 380 nm. The COMSOL simulations also suggest that the ~ 5 nm tuning we observe corresponds to < 10 nm change in the thickness of the $\mu$–disk.

To summarize, we demonstrated a controllable tuning of the WGMs of InGaN/GaN $\mu$–disk cavities. Our results indicate that the tuning occurs due to *in-situ* oxidation of the disk surface and oxide removal by the water. Analysis of the band diagram of the structure and simulations of the mode shifts are in good agreement with our experimental data. Fine tuning of the cavity modes is accessible at low excitation powers, which can be beneficial for tuning PCCs. Tuning availability in III-nitride $\mu$–disk systems would propel the use of III-nitrides in studying cQED systems and development high performance optoelectronic devices.

**Acknowledgments**

The authors thank Andrew Magyar for the help with the cell design. This work was supported by the NSF Materials World Network (Grant No: 1008480) and by the EPSRC under grant number EP/H047816/1.

References

1. K. J. Vahala, Nature **424**, 839 (2003).
2. A. M. Armani, R. P. Kulkarni, S. E. Fraser, R. C. Flagan and K. J. Vahala, Science **317**, 783 (2007).
3. P. Michler, A. Kiraz, L. D. Zhang, C. Becher, E. Hu and A. Imamoglu, Appl. Phys. Lett. **77**, 184 (2000).
4. H. Matsubara, S. Yoshimoto, H. Saito, J. L. Yue, Y. Tanaka and S. Noda, Science **319**, 445 (2008).
5. R. Butte and N. Grandjean, Semicond. Sci. Technol. **26**, 014030 (2011).
6. Y. Arakawa, Ieee Journal of Selected Topics in Quantum Electronics **8**, 823 (2002).
7. E. D. Haberer, R. Sharma, C. Meier, A. R. Stonas, S. Nakamura, S. P. DenBaars and E. L. Hu, Appl. Phys. Lett. **85**, 5179 (2004).
8. Y. S. Choi, K. Hennessy, R. Sharma, E. Haberer, Y. Gao, S. P. DenBaars, S. Nakamura, E. L. Hu and C. Meier, Appl. Phys. Lett. **87**, 243101 (2005).
9. D. Simeonov, E. Feltin, A. Altoukhov, A. Castiglia, J. F. Carlin, R. Butte and N. Grandjean, Appl. Phys. Lett. **92**, 171102 (2008).
10. R. Butte, J. F. Carlin, E. Feltin, M. Gonschorek, S. Nicolay, G. Christmann, D. Simeonov, A. Castiglia, J. Dorsaz, H. J. Buehlmann, S. Christopoulos, G. B. H. von Hoegersthal, A. J. D. Grundy, M. Mosca, C. Pinquier, M. A. Py, F. Demangeot, J. Frandon, P. G. Lagoudakis, J. J. Baumberg and N. Grandjean, Journal of Physics D-Applied Physics **40**, 6328 (2007).
11. S. Mosor, J. Hendrickson, B. C. Richards, J. Sweet, G. Khitrova, H. M. Gibbs, T. Yoshie, A. Scherer, O. B. Shchekin and D. G. Deppe, Appl. Phys. Lett. **87**, 141105 (2005).
12. A. Faraon, D. Englund, D. Bulla, B. Luther-Davies, B. J. Eggleton, N. Stoltz, P. Petroff and J. Vuckovic, Appl. Phys. Lett. **92**, 043123 (2008).
13. D. Sridharan, E. Waks, G. Solomon and J. T. Fourkas, Appl. Phys. Lett. **96**, 153303 (2010).
14. K. Hennessy, A. Badolato, A. Tamboli, P. M. Petroff, E. Hu, M. Atature, J. Dreiser and A. Imamoglu, Appl. Phys. Lett. **87**, 021108 (2005).
15. J. P. Reithmaier, G. Sek, A. Loffler, C. Hofmann, S. Kuhn, S. Reitzenstein, L. V. Keldysh, V. D. Kulakovskii, T. L. Reinecke and A. Forchel, Nature **432**, 197 (2004).
16. A. Faraon and J. Vuckovic, Appl. Phys. Lett. **95**, 043102 (2009).
17. F. Rol, In preparation (2011).
18. R. A. Oliver, G. A. D. Briggs, M. J. Kappers, C. J. Humphreys, S. Yasin, J. H. Rice, J. D. Smith and R. A. Taylor, Appl. Phys. Lett. **83**, 755 (2003).
19. M. S. Minsky, M. White and E. L. Hu, Appl. Phys. Lett. **68**, 1531 (1996).
20. J. W. Seo, C. S. Oh, H. S. Jeong, J. W. Yang, K. Y. Lim, C. J. Yoon and H. J. Lee, Appl. Phys. Lett. **81**, 1029 (2002).

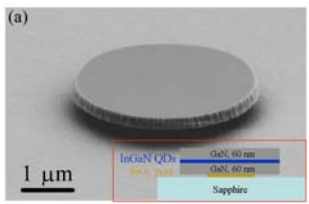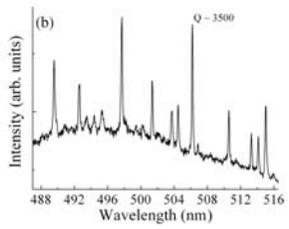